\documentclass[manuscript,screen]{acmart}

\usepackage{textcomp}
\usepackage{xcolor}
\usepackage{multirow}
\usepackage{comment}
\usepackage{booktabs}
\usepackage{colortbl}
\usepackage{hhline}
\usepackage{rotating}
\usepackage{verbatim}
\usepackage{makecell}
\usepackage[ruled,vlined]{algorithm2e}
\usepackage{pifont}
\usepackage{enumitem}
\usepackage[nameinlink,noabbrev]{cleveref}
\usepackage{xurl}
\usepackage{hyperref}
\AtBeginDocument{%
  }

\begin{document}

\title{Synopticon: Consensus-Based Cheating Detection System for Competitive Games}

\author{Jeuk Kang}
\email{gangjeuk@korea.ac.kr}
\orcid{0009-0007-7063-6716}
\affiliation{%
  \institution{School of Cybersecurity, Korea University}
  \city{Seoul}
  \country{Korea}
}

\author{Jungheum Park}
\email{jungheumpark@korea.ac.kr}
\orcid{0000-0001-7796-7699}
\affiliation{%
  \institution{School of Cybersecurity, Korea University}
  \city{Seoul}
  \country{Korea}}

\renewcommand{\shortauthors}{Trovato et al.}
\renewcommand{\algorithmautorefname}{Algorithm}
\renewcommand{\sectionautorefname}{Section}
\renewcommand{\subsectionautorefname}{Section}
\renewcommand{\subsubsectionautorefname}{Section}
\setlength{\algomargin}{1em} 

\begin{abstract}
Cheating in online games has serious impacts on the game industry in various ways. Although extensive research has been conducted in this field, most studies have focused on a specific genre, the Massively Multiplayer Online Role-Playing Game (MMORPG). 
On the other hand, competitive genres---such as Multiplayer Online Battle Arena (MOBA), First Person Shooter (FPS), Real Time Strategy (RTS), Action---have received less attention compared to MMORPG, as both, the difficulty in detecting cheating users and the demand for sophisticated techniques and data, make them less attractive. Rather than take the harder path, game companies, which make competitive games, have adopted an anti-cheat solution and have urged users to install it before playing their games. However, the anti-cheat solution demands kernel-level permission, and its operations are concealed from users. Due to its features, it has been used for attacks by hackers taking advantage of its high authority, and users are posing questions about their privacy. In this paper, we introduce \texttt{SYNOPTICON}, a novel cheating detection system. \texttt{SYNOPTICON} detects cheating users by using normal users' consensus, which is achieved by users' votes. \texttt{SYNOPTICON} consists of a detection method on client-side and a voting system on server-side. When abnormal behavior of cheating users is detected by a client, it casts a vote to the server, which then distinguishes between cheating users and normal users based on the aggregated votes and the consensus achieved by the votes. 
This design allows \texttt{SYNOPTICON} to address the key challenges faced by cheating detection systems. \texttt{SYNOPTICON} has been tested in both a simulation and a real-world FPS game. With the simulation, we verify the capabilities and requirements of \texttt{SYNOPTICON}, and, with the real-world game, we evaluate its effectiveness in identifying cheating users. As a result, \texttt{SYNOPTICON} successfully detected cheating users. In addition, we demonstrate its applicability and sustainability in managing online games through a real-like operational environment developed using public datasets.
\end{abstract}



\maketitle

\section{Introduction}\label{sec:introduction}
Cheating in computer games is often perceived as acts of \textit{gaining advantages} by abusing a game system (\cite{consalvo2009cheating}). In the early stage of the game industry, cheating emerged from single-playing games and was used for improving a player's personal satisfaction by manipulating the difficulty or improving the handiness of a game (\cite{passmore2020cheating}).
As time passed, online multi-playing games appeared, and soon, became the mainstream of the game industry. Cheating was also on the stream, and its characteristics have changed reflecting the changes in the industry. Most recently, cheating is interfering in interactions between online players (or users) and is damaging the game industry (\cite{bbc_hacker}).

To solve these, numerous studies on anti-cheat methods, which detect and counter cheating, have been conducted, yet the majority of these studies have concentrated on one specific genre of online game named Massively Multiplayer Online Role-Playing Game (MMORPG). In addition, much of the research has targeted cheating aimed at financial gain. Survey papers (\cite{woo2012survey, han2022cheating, kotkov2018gaming}), that went over anti-cheat methods indicate that most research has focused on MMORPGs, while relatively few studies addressing competitive genres---such as Multiplayer Online Battle Arena (MOBA), First Person Shooter (FPS), Real Time Strategy (RTS), Action. This disparity is largely due to the fact that the \textbf{\textit{means}} and \textbf{\textit{objectives}} of cheating vary significantly depending on the \textit{ultimate goal} of each genre. In a single-player game, the goal is to complete a game or simply to enjoy a game. In MMORPG, it is usually set to the growth of players' characters, in line with the Role-Playing nature of the genre. In competitive genre, it is set to a victory in competition between players. Such differences in the \textit{goals} make distinct differences to cheating as well. For example, in competitive genre, cheating operates as a supporter and guarantees a winning by propping up the player's performance and by providing additional information that cannot be acquired in a normal way. Further discussion on the \textbf{\textit{means}} and \textbf{\textit{objectives}} of cheating across various game genres will be provided in \autoref{sec:background}.

However, there are several challenges with the existing cheating detection methods that have been proposed for the competitive genre. 

\begin{description}[leftmargin=*]
   
  \item [Ch1: Lack of consideration for liveness] 
  Our investigation in \autoref{sec:background} reveals that most of the anti-cheat methods applied in MMORPG rely on filtering out cheating users by analyzing long-term logs on game servers. However, such approaches are inadequate for competitive genre; mainly because, many competitive games are played on a match-basis. In these games, a matching system sets two groups (or teams, e.g., \textit{N vs N}) of users, and matches between two groups are played. In contrast, users in MMORPG are playing a game in a virtual world with a continuous timeline. Therefore, cheating detection in competitive genre should be rapid and performed within a single match. In short, the method must satisfy the liveness, ensuring quick detection and high accuracy during a match. Moreover, the liveness is crucial for an aspect of user experiences; because, in competitive genre, cheating directly influences normal users and harms their impression on a game by spoiling user experiences and giving a sense of losing during the whole match time.
   
  \item [Ch2: High overhead] 
  The introduction of cheating detection systems inevitably incurs overhead on servers or clients, and many methods applied in MMORPG pass on their costs to servers. However, this approach is less feasible for competitive genre; because, it requires more sophisticated datasets and techniques, both increasing the cost. For instance, Valve, an American game company, implemented a cheating detection system on their servers to filter out cheating users in one of their games named `Counter-Strike: Global Offensive'. Consequently, they had to use the server machines with 3,456-cores for one day (\cite{Robocalypse_Now}), and it is expected that Valve had to spend a few million dollars at least (\cite{published_valve_2018}) to maintain the server. Some game companies tried to pass on their costs to users by involving users as judges in the ruling process (\cite{Tassi, Ask_Riot_2018, Counter-Strike}). However, their attempt failed because of biases in human judgments and the inefficiency of the system itself (\cite{Ask_Riot_2018}).

  \item [Ch3: Low applicability] 
  Client-side methods, on the other hand, have suffered from their low applicability. Recent studies already reached the state-of-the-art in detection rate of cheating users by using a mix of sophisticated machine learning methods and detailed datasets; however, regardless of its accuracy, it is still remained vulnerable to tampering by clients. Several papers adopted the Trusted Execution Environment (TEE) (\cite{Choi_Ko, Park_Ahmad_Lee_2020, jeon2021tzmon}) to prevent manipulations by clients. However, the use of TEE is yet impractical---due to the limited number of CPU models that support TEE in the personal computer market, and the inaccessibility of TEE for game companies in the mobile market---which make it hard for the companies to apply TEE on their games.

  \item [Ch4: Low sustainability] 
  Cheating detection systems need constant updates and re-adjustment, as parameters and variables in games, which researchers used for the optimization of their methods, can change throughout the updates on games, such as re-balancing patches. In addition, the systems also have to deal with new cheating methods. Otherwise, they will be outdated soon. Recent research for MMORPG has just started to address the importance of sustainability and has begun to discuss the operational aspects of their methods (\cite{Lee_Woo_Kim_Mohaisen_Kim_2016, Tao_Xu_Gong_Li_Fan_Zhao_2018, Xu_Luo_Tao_Fan_Zhao_Lu_2020}). In competitive genre, however, it hasn't been discussed yet.
  
\end{description}

Giant game companies (e.g., EA, Epic Games, Riot of Tencent, Activision Blizzard of Microsoft) handled these challenges with client-side anti-cheat solutions (\cite{call_of_duty_ricochet, riot_vanguard_dev_2024, Steam_anti_cheat, ea_Arts_2022}). Many of them are forcing users to install it before playing their games. For instance, Epic Games, known for its famous game `Fortnite' with 400 million global users (\cite{aws_fortnite_case_study}), mandates the installation of their anti-cheat solution (\cite{fortnite_anti_cheat}). These solutions run at kernel-level and operate by monitoring external programs' interference in a game-client, a program that runs a game. Unlike other methods of previous studies, they do not rely on sophisticated machine learning techniques and detailed datasets (\textbf{Ch3}). They operate on client-side (\textbf{Ch2}) and run in real-time (\textbf{Ch1}). Finally, they are continually updated and managed by fixing security loopholes and responding to evolving cheating methods (\textbf{Ch4}).

However, the cost of anti-cheat solutions, which may seem reasonable solutions for game companies, are paid by clients' privacy. Vulnerabilities in these (\cite{CVE-2005-0295,CVE-2020-36603,CVE-2022-27095}) have been used by attackers (\cite{genchin_ransomware}), and the high privileges granted to the solutions---intended to assure their unconstrained operations---is then offering advantages to attackers (\cite{maario2021redefining,dorner2024if}). Furthermore, because of its unconstrained operations, it has been accused of illegal gathering of information. Game companies are trying to refute these accusations against them. For instance, Riot Games---which developed its own anti-cheat solution named \textit{Vanguard} (\cite{What_is_Vanguard?_2024}) and integrated the solution with its games---has offered \$100,000 bug-bounty reward to demonstrate its commitment on security (\cite{Riot_Games_Vulnerability_Disclosure_Policy}). Likewise, others have made efforts to prove their solutions are both safe and privacy-preserving (\cite{riot_vanguard_dev_2024, call_of_duty_ricochet,ea_Arts_2022,Park_2020}). Nevertheless, despite these efforts, anti-cheat solutions are still yet bypass-able (\cite{Black_Hat_bypassing_anticheat,Guigo_John}), and threats against the solutions are ongoing problems.

In this paper, we introduce \verb|SYNOPTICON|, a novel cheating detection system based on the consensus of normal users. Inspired by Jeremy Bentham's concept of `Panopticon' (\cite{bentham1791panopticon}), \verb|SYNOPTICON| leverage mutual surveillance, where users evaluate each other, and \verb|SYNOPTICON| finds bad guys by using a consensus among users achieved by the mutual evaluations. By utilizing the collective judgment of users as the \textbf{Ground truth}, \verb|SYNOPTICON| effectively differentiates between cheating users and normal users. \verb|SYNOPTICON| can be implemented on client-side (\textbf{Ch2}) and overcomes the inherent limitations of client-side methods (\textbf{Ch3})---without the need for additional technologies, such as TEE. We evaluate \verb|SYNOPTICON| using both simulations and a real-world dataset. Our evaluation demonstrates that \verb|SYNOPTICON| can detect cheating users with high accuracy and speed (\textbf{Ch1}).
Furthermore, we explore how \verb|SYNOPTICON| can be applied in terms of game service management and operating (\textbf{Ch4}). 

Finally, a proof-of-concept version of \verb|SYNOPTICON| and the associated datasets are \href{http://anonymous.4open.science/r/Gynopticon-91C6}{publicly available}, allowing this work to be reproduced.

\section{Background} \label{sec:background}

\begin{table*}[ht]
\centering
\caption{Typical cheating behaviors in online multi-playing games}
\label{tab1:scope-of-paper}
\begin{tabular}{llll} 
\toprule
\textbf{Genre}               & \textbf{Subgenre}                                                               & \textbf{Cheating}                                                       & \textbf{Description}                                                                                                                                                  \\ 
\midrule
Common                       & -                                                                               & Exploits                                                                & \begin{tabular}[c]{@{}l@{}}Using of bugs and manipulation of certain features on games (e.g., \\speed hack that allows rapid move, ghosting that is being invisible)\end{tabular}  \\ 
\midrule
Role-Playing                 & MMORPG                                                                          & Game bots                                                               & Operate automated bots and gather virtual property or achievement                                                                                                      \\ 
\midrule
\multirow{2}{*}{Competitive} & \multirow{2}{*}{\begin{tabular}[c]{@{}l@{}}FPS\\MOBA\\RTS\\Action\end{tabular}} & Scripting                                                               & \begin{tabular}[c]{@{}l@{}}Automate certain behavior or action and assure fast sequential\\inputs or rapid response time 
 (e.g., aimbots on FPS)\end{tabular}          \\ 
\cmidrule{3-4}
                             &                                                                                 & \begin{tabular}[c]{@{}l@{}}Extra-Sensory\\Perception (ESP)\end{tabular} & \begin{tabular}[c]{@{}l@{}}Disclose hidden information, such as, location of other users\\ behind wall which should not be displayed in normal\end{tabular}           \\
\bottomrule
\end{tabular}
\end{table*}

\subsection{Type of game cheating} \label{subsec:background-various-cheat}

As mentioned in \autoref{sec:introduction}, cheatings are typically employed to gain \textit{benefits}, that cannot be obtained in a normal way, and the nature of these benefits varies by the genre and \textit{ultimate goal} of a game. If a goal is set for the growth of characters, cheating can be used to earn game money to purchase new gadgets (e.g., new swords and shields) and to upgrade the ability of their character (e.g., power and speed). When it is a victory in competition (\cite{lee2021some}), cheating can serve as an assistant by encouraging cheating users to perform beyond their innate abilities. We have classified types of game cheating according to game genres and cheating methods as listed on \autoref{tab1:scope-of-paper}.

\subsection{Existing cheating detection methods} \label{subsec:background-various-anti-cheat}

\begin{table}[ht!]
\caption{Number of existing research papers on MMORPG}
\label{tab2:previous_study}
\centering
\begin{tabular}{lrrrrr} 
\toprule
\textbf{Detection side} & \multicolumn{4}{c}{\textbf{ Used Method }}  \\ 
\midrule
                                            & STAT & ML & SIMI & NET                      \\ 
\cmidrule{2-5}
Server-side                                 & 10   & 21 & 3    & 8                        \\
Client-side                                 & -    & 3  & -    & -                        \\
Network-side                                & 4    & -  & -    & -                        \\
\bottomrule
\end{tabular}
 \vspace{1ex}
 
 {\small \centering STAT: Statistics analysis, ML: Machine learning\\SIMI: Similarity analysis, NET: Network based\par}
\end{table}

Cheating detection methods can be categorized into two primary types depending on where they are applied: client-side and server-side. We reviewed survey papers (\cite{woo2012survey,han2022cheating,kotkov2018gaming}) that investigated cheating detection methods targeting MMORPG genre, which has been researched most. Detection methods in the papers were classified according to the type of the methods---whether client-side or server-side---as well as the specific techniques employed. 
As shown in \autoref{tab2:previous_study}, most research in MMORPG genre has adopted server-side, and its methods are mainly about analysis of server-side logs. This approach works well in detecting \textit{gamebots}, as their behavior patterns differ significantly from those of normal users in a long-term view. However, it is less effective for detecting scripting cheating in competitive genre, where games are typically played on match-basis.

\subsection{Research scope}

Before explaining \verb|SYNOPTICON|, we clarify the scope of its applicability. From the top of the \autoref{tab1:scope-of-paper}, first, MMORPG and \textit{gamebots} are outside the scope of this study. Regarding Role-Playing, extensive research has already been conducted on \textit{gamebot} detection, and several state-of-the-art methods have been developed (\cite{mitterhofer2009server,Lee_Woo_Kim_Mohaisen_Kim_2016,Tao_Xu_Gong_Li_Fan_Zhao_2018, Xu_Luo_Tao_Fan_Zhao_Lu_2020}). Second, our focus is on the competitive genre and the detection of scripting, where the advantages of \verb|SYNOPTICON| can be used most effectively. Meanwhile, other cheatings on competitive genre are not addressed in the paper. In the case of \textit{exploits}, it can easily be defended in modern online games, as most of the states and features of games are calculated and managed on server-side, not client-side. Epic Games---which developed the \textit{Unreal Engine}, one of two central pillars of 3D game engine---explains to developers that ``cheats allowing players to fly, teleport, or move extremely fast can usually be completely prevented at the game architecture level by giving game servers authority over player positions and movement (\cite{epic_games_dev_anti_cheat}).'' 

Detecting \textit{gamebots} has already been researched in MMORPG and it is not a primary concern in competitive genre. In the case of Extra-Sensory Perception (\textit{ESP}) cheats, game companies have proposed solutions in various ways; from a policy-level solution that implements honeypots in a game-client (\cite{Cheaters_Will_Never_Be_Welcome_in_Dota}), to a technical solution that limits a game server to send needless information that could be misused for providing additional information (\cite{pubg_anti_cheat_introd}). 

\verb|SYNOPTICON| can be integrated with various cheating detection methods previously studied---such as statistics analysis, similarity analysis and machine learning---but it operates under the assumption that all users can evaluate each other mutually. For instance, vision-based methods, which detects cheating users by evaluating the information on the user's screen, fall outside the scope, as they rely solely on data from a single user. Lastly, while \verb|SYNOPTICON| can be applied across all competitive genres, for simplicity and consistency in terminology, the explanation in \autoref{sec:Methodoloy} will focus on its application to \textit{aimbots}, a common type of scripting in FPS genre.

\section{Synopticon: consensus-based cheating detection system for competitive games}
\label{sec:Methodoloy}

\subsection{Overview}
\label{sec:methodoloy-overview}

\begin{figure*}[htbp]
\includegraphics[width=\textwidth]{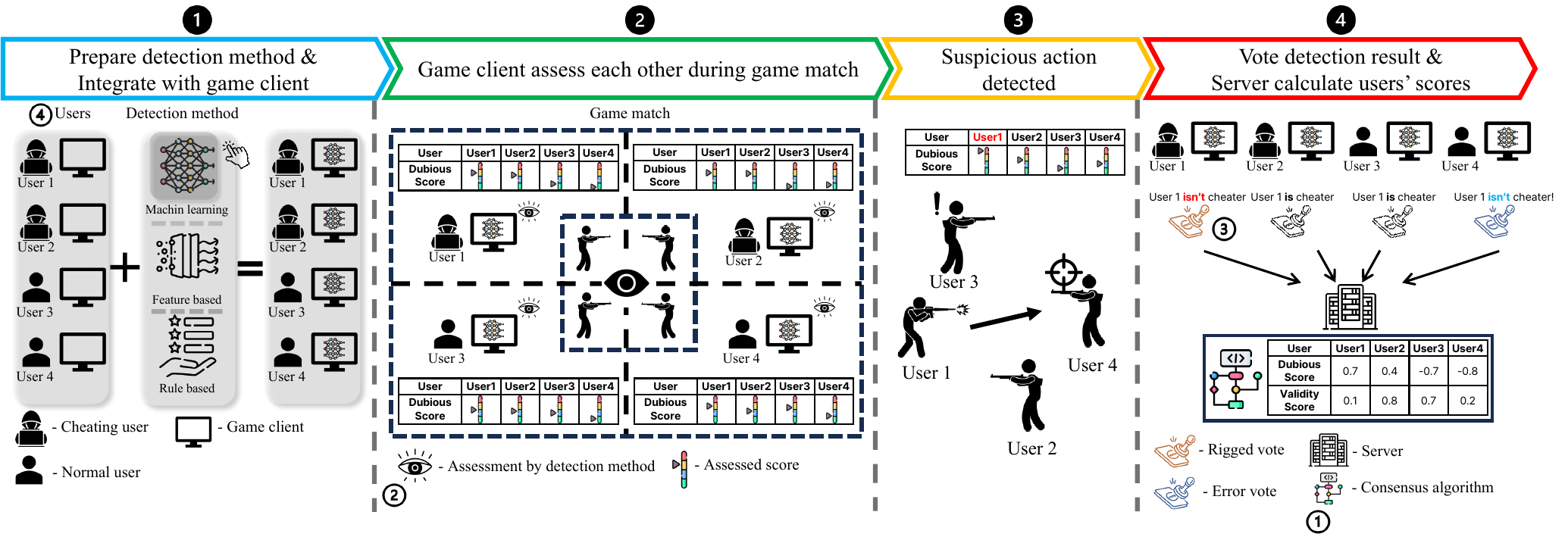}
\captionof{figure}{Overall design of \texttt{SYNOPTICON}: the design and concept is described in \autoref{sec:methodoloy-overview}---\ding{202},\ding{203},\ding{204},\ding{205}, and the basic requirements is discussed in \autoref{subsec:simulation_overview}---\ding{192},\ding{193},\ding{194},\ding{195}}
\label{fig_gynopticon_overall}
\Description[Overall design of SYNOPTICON]{the design and concept is described in \autoref{sec:methodoloy-overview}---\ding{202},\ding{203},\ding{204},\ding{205}, and the basic requirements is discussed in \autoref{subsec:simulation_overview}---\ding{192},\ding{193},\ding{194},\ding{195}}
\end{figure*}

The \verb|SYNOPTICON| assumes a general structure of online competitive games, that comprises users who play a game and a game server that supervises a game. To play a match, users request match-making from the server, and the server sets a match which is generally made up of competitions between two groups of users (e.g., \textit{N vs N}). During the match, the server receives data from client-side (e.g., keyboard-inputs for movement and mouse-inputs for gunfire), and the server continuously calculates and updates the state of the game. Then, the server broadcasts the updated state to users, and the state is rendered by a game-client at client-side (e.g., movement of characters and graphical effects of gunfire). During this process, \verb|SYNOPTICON| works as shown in \autoref{fig_gynopticon_overall}.
 
\begin{description}[style=sameline,leftmargin=*]

    \item[\ding{202}] Detection method is a method for detecting cheating users. For instance, one of many client-side methods suggested in previous studies can be used as a detection method, regardless of its approach---such as machine learning, statistics analysis and rule-based detection---as long as it can evaluate user's behavior by using data received from the server. The detection method is implemented in a game-client and monitors users' behavior during a match using the received data. 

   \item[\ding{203}] Information from the server contains not only one player's data but also other players' data nearby, so the game-client can render the states of a game. For example, the game-client receives the states---such as character's movement by keyboard-inputs and gunfire by mouse-inputs---from the server and renders all characters' movements nearby, and graphical effects of flames and explosion of the gunfire in battlefield. By using these sets of data, the detection method in each user's game-client can evaluates each other's behavior during a match.

   \item[\ding{204}] If any detection method in each user detects an abnormal (when a suspicious score exceeds a threshold), a \textit{vote} begins. Each user casts their evaluation results as a ballot to the server. One ballot contains information about whether one user is a cheating user or a normal user.

   \item[\ding{205}] Based on the users' ballots, the server uses a consensus algorithm to calculate each user's score and differentiates between cheating users and normal users after a match.   
 \end{description}
 
To sum up, \verb|SYNOPTICON| consists of two main components: the detection method on client-side and the consensus algorithm on server-side. The details of each will be described in the following sub-sections.

\subsection{(Client-side) Voting initiator with appropriate detection methods}

\verb|SYNOPTICON| assumes two types of users, \textit{cheating} users and \textit{normal} users. All users play a game using a game-client that contains appropriate detection methods. Therefore, \textit{vote} can occur when the detection method identifies abnormal actions of a cheating user. If the method successfully distinguishes between cheating and normal users, each user will cast \textit{True} for a cheating user and \textit{False} for a normal user to the server.

Besides accurate evaluations, there may be two additional scenarios of mis-estimation. First, users might cast an \textbf{\textit{error vote}} when the evaluation of the method is wrong. Secondly, cheating users can manipulate their ballot and send a \textbf{\textit{rigged vote}} to the server, as the client-side methods are vulnerable to tampering by cheating users. However, the \textbf{\textit{rigged vote}} by normal users is not considered.

\subsection{(Server-side) Cheater discriminator with a proposed consensus algorithm}

\begin{algorithm}[ht!]
\SetKwData{True}{True}
\SetKwData{False}{False}
\SetKwFunction{Count}{count}
\SetKwFunction{Enque}{enque}
\SetKwFunction{Deque}{deque}
\SetKwFunction{Pop}{pop}
\SetKwFunction{Length}{length}

\SetKwInput{GT}{\textit{GT}}
\SetKwInput{Users}{\textit{Users}}
\SetKwInput{VadW}{$W_v$}
\SetKwInput{HisW}{$W_h$}
\SetKwInput{History}{$Q_N$}
\SetKwInput{Dubious}{$D_N$}
\SetKwInput{Validity}{$V_N$}
\SetKwInput{UserV}{$Vote_{target}(N)$}
\SetKwFunction{FMain}{score\_evaluation}
\SetKwFunction{FDoVote}{do\_vote}
\SetKwProg{Fn}{Function}{:}{}

 \GT {Ground truth}
 \Users {List of users}
 \VadW {Weight for Validity}
 \HisW {Weight for History}
 \History {Queue for voting record of user\textit{N}} 
 \Dubious {Dubious score of user\textit{N}}
 \Validity {Validity score of user\textit{N} $(0\le V_N\le 1)$}
 \UserV {A vote of user\textit{N} for \textit{target}}
 
  \Fn{\FMain{$target\_user$, \textbf{GT}}}{
    \ForEach{$user$ in the \textbf{Users}}{
        t\_cnt $\leftarrow Q_{user}.$\Count{\True}\;
        f\_cnt $\leftarrow Q_{user}.$\Count{\False}\;
        hist\_record $\leftarrow$ t\_cnt - f\_cnt\;
        \lIf{$Q_{user}.\Length{}>10$}{$Q_{user}.$\Deque{}}
        \uIf{$Vote_{target\_user}(user)\ne\textbf{GT}$}{user\_lied $\leftarrow$\True\;} \uElse{user\_lied $\leftarrow$\False\;}
        
        \uIf{user\_lied is \True}{
            $V_{user}$ $\leftarrow$ $V_{user}-W_v + hist\_record*W_h$\;
            $Q_{user}.$\Enque{\False}\;
        }
        \Else{
            $V_{user}$ $\leftarrow V_{user}+W_v+hist\_record*W_h$\;
            $Q_{user}.$\Enque{\True}\;
        }

        $D_{target\_user}$ $\leftarrow D_{target\_user}+V_{user}$\;
    } 
  }

\Fn{\FDoVote{\textbf{Users}, \textbf{GT}}}{
    \ForEach{$target\_user$ in the \textbf{Users}}{
        \FMain{$target\_user$, \textbf{GT}}
    } 
  }

\caption{\textbf{Validity}, \textbf{Dubious} score evaluation}
\label{algo:censensus_algorithm}

\end{algorithm}

The server evaluates users' scores using a consensus algorithm for each \textit{vote} that occurs during a match. In our design, we introduced two scores, \textbf{Validity} and \textbf{Dubious}, for the evaluation. 

At the beginning of each match, both scores are set to initial state for all users. To clarify, since no information about users is given to the server, all users have the maximum \textbf{Validity} score of 1 and the neutral \textbf{Dubious} score of 0. The server updates both scores for each \textit{vote} based on the \textbf{Ground truth} decided by users' ballots for each \textit{vote}. The server sets the \textbf{Ground truth} according to the majority of ballots, since the server doesn't know who is a cheating user. 
The \textit{vote} assumes a secret vote, in which one's poll is concealed from others. This assumption is reasonable; because, matches between users are randomly set, and it is unlikely that they know each other and cooperate to manipulate the \textit{vote}.

The two scores are made by a proposed consensus algorithm on server-side and its details are on \autoref{algo:censensus_algorithm}. Two factors were considered for the design of the algorithm. First, a trustworthy user affects more on the \textit{vote}. Depending on the \textbf{Validity} score, users with higher scores influence more on the evaluation of the \textbf{Dubious} score, while those with lower scores have less impact. Second, we added a voting history for the evaluation of \textbf{Validity}. It is to take the \textbf{\textit{error vote}} occurred by wrong evaluation into consideration. Users, who cast the vote consistent with the \textbf{Ground truth} frequently, get less penalty on their \textbf{Validity}, even if they vote incorrectly. In contrast, more penalties will be imposed on users who lie often, and, as a result, they will lose their influence on the \textit{vote} rapidly, with their decreasing \textbf{Validity}. 

In the end, \verb|SYNOPTICON| considers users with high \textbf{Dubious} score as cheating users. This design allows \verb|SYNOPTICON| to detect cheating users, while effectively keeping the overhead on server-side low. At the same time, it successfully prevent the influence of \textbf{\textit{rigged vote}} on the \textit{vote} by cheating users who are potential adversaries of client-side methods and are able to tamper, repudiate, and manipulate the results. For example, if cheating users do vote honestly, they have to admit that they are cheaters. If they do \textbf{\textit{rigged vote}} and say lie against the \textbf{Ground truth} to repudiate charges on them, their \textbf{Validity} would decrease and lose their influence on the \textit{vote}. Ultimately, cheating users must either admit to cheating or lie against the \textbf{Ground truth}, which increase their \textbf{Dubious} scores or decrease their \textbf{Validity} scores.

However, as mentioned above, we assumed that an appropriate existing detection method is available, so if the detection method cannot distinguish between cheaters and normal users well enough, \verb|SYNOPTICON| will not work properly. Therefore, the following subsection addresses some additional research questions to determine the detailed operational parameters for the proposed system to work in practice. 

\subsection{Operational parameter testing with simulation} \label{subsec:simulation_overview}

To assess the effectiveness of \verb|SYNOPTICON| as a cheating detection system and to determine the minimum requirements for its successful operation, we conducted a series of experiments with simulation. These experiments were designed to address the following four questions and associated critical factors:

\begin{description}
    \item[\textbf{Q1-1}] Will the consensus algorithm work?
\begin{description}
    \item[\ding{192}] Final dubious score of users after each match
\end{description}
    \item[\textbf{Q1-2}] How much of \textit{model acc} is needed for accurate detection?
\begin{description}
    \item[\ding{193}] Least accuracy of detection method for good-enough performance
\end{description}
\item[\textbf{Q2-1}] How powerfully can cheating users influence the \textit{vote}?
\begin{description}
    \item[\ding{194}] Influence of \textbf{\textit{rigged vote}} by cheating users
\end{description}
    \item[\textbf{Q2-2}] How many cheating users can be endured?
\begin{description}
    \item[\ding{195}] Maximum ratio of cheating users that can be handled
\end{description}
\end{description}

The simulation comprises multiple users and a server, and there are four controllable variables as follows: \textit{vote count}, the number of votes occurred during one match. \textit{benign num}, the number of normal users in a match. \textit{cheater num}, the number of cheating users in a match. \textit{model acc}, the accuracy of the detection method. We performed two experiments by changing the variables to answers \textbf{Q1-1} and \textbf{Q1-2} respectively. In addition, for \textbf{Q2-1} and \textbf{Q2-2}, we reflected the \textbf{\textit{rigged vote}} to the simulation. For that, we supposed that cheating users can behave tactically. Hence, we carried out each experiment on three different conditions: \texttt{Without liar} that no users poll the \textbf{\textit{rigged vote}}. \texttt{Random liar} that cheating users can manipulate the ballot and do \textbf{\textit{rigged vote}} for one-half of their ballot---for left one-half, however, they simply poll honestly. \texttt{Tactical liar} that a cheating user always votes on not cheater (\textit{False}) for oneself. Every match in the simulation is independent mutually; in other words, all users' scores are reset to the initial state for every match.

\subsubsection{Answer to Q1-1 and Q2-1} 
\label{subsec:existence_of_liar}

\begin{figure}[h]
\includegraphics[width=\columnwidth]{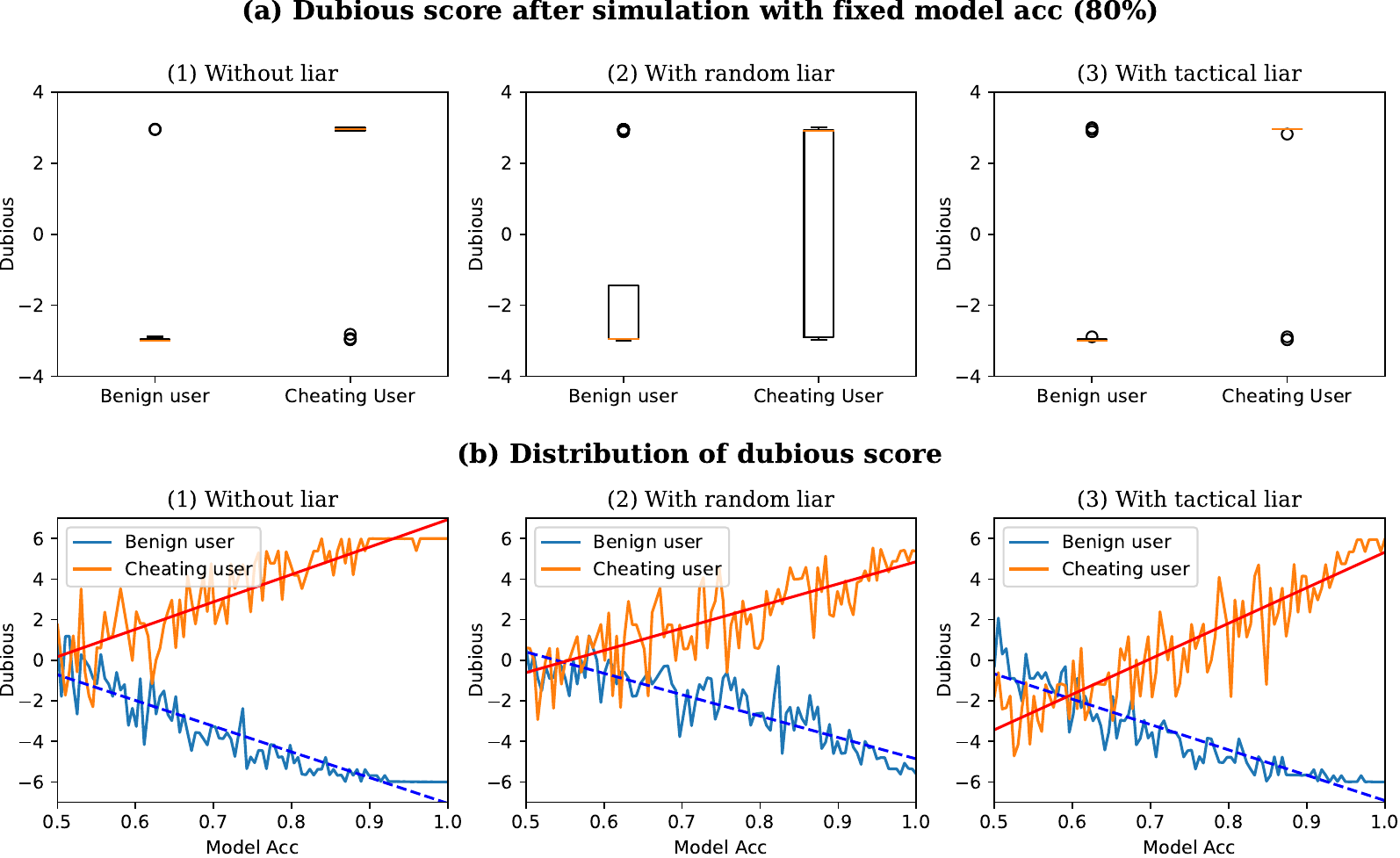}
\captionof{figure}{Results of the first simulation. The proposed consensus algorithm was examined. \textbf{(a)} demonstrates the \textbf{Dubious} scores of normal users and cheating users at fixed 80\% of \textit{model acc}. \textbf{(b)} demonstrates the scores of cheating and normal users at different \textit{model acc} from 50\% to 100\%}
\label{fig_scatter}
\Description[Results of the first simulation]{The proposed consensus algorithm was examined. (a) demonstrates the \textbf{Dubious} scores of normal users and cheating users at fixed 80\% of model acc. (b) demonstrates the scores of cheating and normal users at different model acc from 50\% to 100\%}

\end{figure}

In the first simulation, we verified that our consensus algorithm actually works by recording the final \textbf{Dubious} score of users after each match. The simulation supposed two normal users and one cheating user as players, and only one \textit{vote count} in a match. The \textit{model acc} was set to 80\%. The reason for the small \textit{vote count} and small number of users is to reflect the influence of the \textit{model acc} and to reduce the compensations from the voting history record in our consensus algorithm. Because, as more voters participate in a vote and as more ballots are cast, the results will be stabilized, yet we want harsher condition for the simulation.

After several matches, we analyzed the median value of the final \textbf{Dubious} scores after matches by using the box plots. In addition, to examine the correlation between \textit{model acc} and the final \textbf{Dubious} score, we observed shifts in the median scores of each two groups by changing the \textit{model acc}. The result on \autoref{fig_scatter}-(a) proves that scores between two groups are distinguishable, confirming that our consensus algorithm functioned as intended. \autoref{fig_scatter}-(b) demonstrates that the tactics of cheating users had an effect. Especially on -(b)-(2) and -(b)-(3), a gap between two groups are more narrow compared to -(b)-(1). Moreover, the results underscores the importance of \textit{model acc}, as the gap narrowed sharply in accordance with the decreasing of \textit{model acc}

\subsubsection{Answer to Q1-2 and Q2-2}
\begin{figure}[h]
\includegraphics[width=\columnwidth]{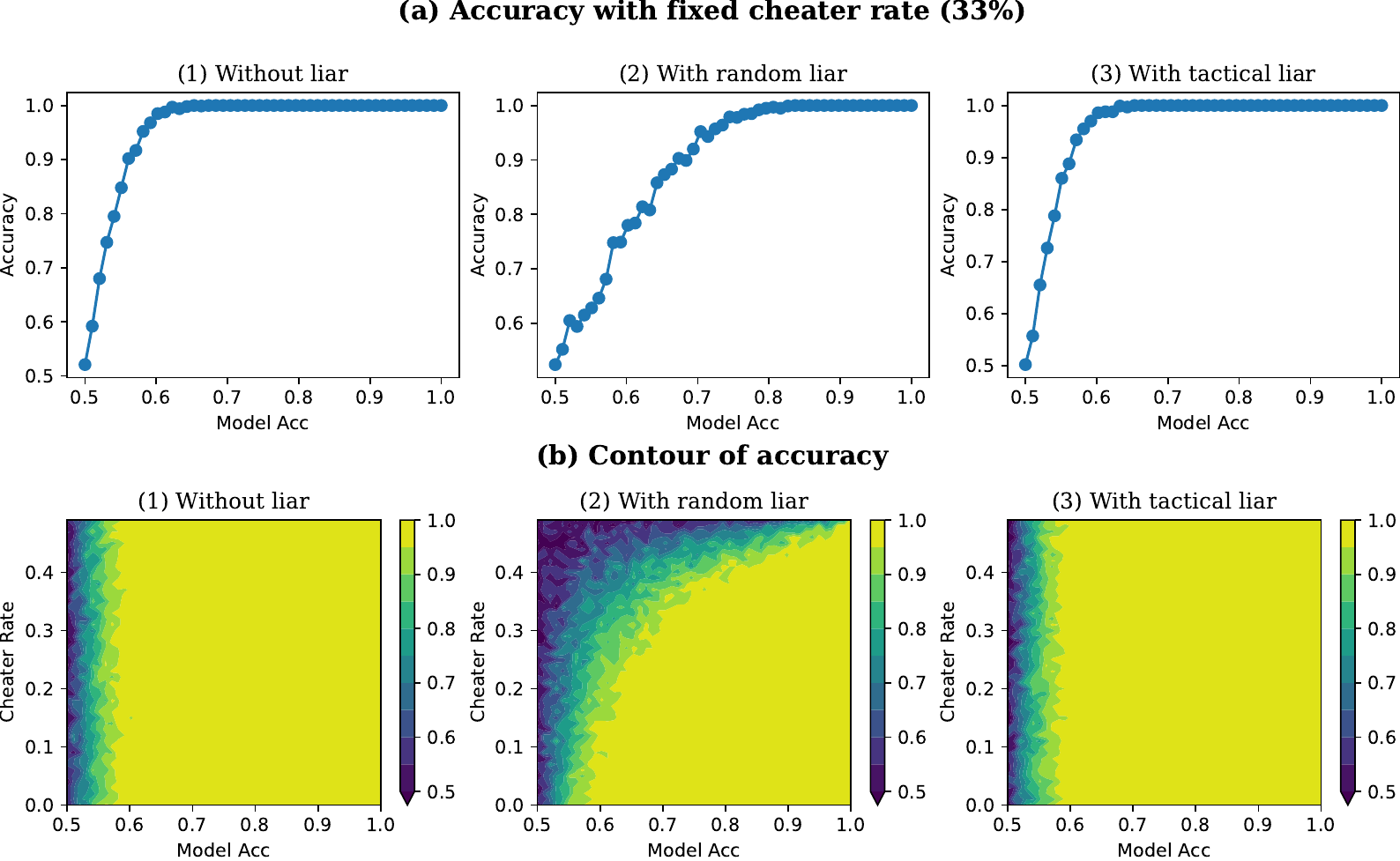}
\captionof{figure}{Results of the second simulation. The minimum \textit{model acc} was examined. \textbf{(a)} demonstrates the needed accuracy at a fixed cheater rate of 33\%. \textbf{(b)} demonstrates the changing accuracy at different cheater rates from 0\% to 50\%}
\label{fig_Contour}
\Description[Results of the second simulation.]{The minimum model acc was examined. (a) demonstrates the needed accuracy at a fixed cheater rate of 33\%. (b) demonstrates the changing accuracy at different cheater rates from 0\% to 50\%}
\end{figure}

In the second simulation, we observed how much of the \textit{model acc} is required to achieve the high accuracy of classification. In order to answer the question, we evaluated the \textit{Accuracy} of the classification by considering users with \textbf{Dubious} scores below zero as normal users and those above zero as cheating users. In this simulation environment, we increased the total number of users to 100, in order to adjust the ratio between cheating and normal users.

Simulation results on \autoref{fig_Contour}-(a) indicate that an approximate 80\% of the \textit{model acc} is required to ensure the roughly 100\% of \textit{Accuracy}, when the ratio of cheating users fixed on 33\%. Furthermore, we also examined how the \textit{Accuracy} showed on \autoref{fig_Contour}-(a) changes depending on the ratio of cheating users, from 0\% to 50\%. The results on \autoref{fig_Contour}-(b) illustrates that if \textit{model acc} satisfies more than 80\%, \verb|SYNOPTICON| can assure nearly 100\% of \textit{Accuracy} under the condition of roughly 40\% of cheating users. On the other hand, it is hard to imagine the case that cheating users account for more than 40\% among all users. For example, in the case of \textit{(5 vs 5)} competitive games like `Counter-Strike: Global Offensive', even if only 7\% of the total users are using cheats, there is approximately a 50\% chance that there will be at least one cheating user in the match (\cite{Robocalypse_Now}), proving that such high rate is unrealistic. If it exists, otherwise, it is unlikely to consider that such games are properly managed.

In addition, results on \autoref{fig_Contour}-(a)-(3) and \autoref{fig_Contour}-(b)-(3) suggest that the tactics of liar worked less effectively than in the first simulation, and it shows that increased number of voters and ballots effects the stability of results. However, the tactic of \texttt{Random liar} still worked, and the results on \autoref{fig_Contour}-(b)-(2) indicates that \verb|SYNOPTICON| becomes ineffective when 50\% of users are cheating users; because, in such a situation, the \textbf{Ground truth} can be manipulated by cheating users.

\section{Evaluation} \label{subsec:real_case_study}

In this section, we apply \verb|SYNOPTICON| to a public dataset of an online game named `Counter-Strike: Global Offensive' for the purpose of evaluation.

\subsection{Dataset description}

The dataset was released by \cite{Choi_Ko}. The research focused on client-side detection method for \textit{aimbots}. They employed a variety of users, from professional players to average players, and recorded each player's game-playing logs at client-side. The players played for several matches, and, before each match, a subset of players was asked to use \textit{aimbots}. Since their method was applied on client-side, the log contains the data received from a server. Specifically, the data contains timestamp, player's current position (\texttt{x}, \texttt{y}, \texttt{z}), player's aiming angle (\texttt{ax}, \texttt{ay}, \texttt{az}) and in-game event (\texttt{fire}, \texttt{hit}, \texttt{dead}). 
As a detection method, they employed a Recurrent Neural Network (RNN), one of the machine learning techniques, to differentiate cheating users. Because of supports from \textit{aimbots}, cheating user's actions are different from those of normal user, and the method can detect cheating users by detecting such differences. For example, cheating user's aiming angle may change so rapidly that even professional players can't imitate it.

For evaluation, we preprocessed the dataset and customized our consensus algorithm considering the characteristics of the FPS genre. According to the authors of the dataset (\cite{Choi_Ko}), their detection method reached more than 90\% of accuracy, which meets our minimum requirement of more than 80\% \textit{model acc} for almost 100\% of \textit{Accuracy}.

\subsection{Preprocessing of dataset}
First, the overall timeline of a match has been reconstructed by integrating every user's log and has been divided into small timelines. The smaller timelines have been divided based on a battle between users; because, in most cases, \textit{aimbots} is used for a battle. The slide window of the battle has been sliced considering the timestamps when \texttt{fire} events were recorded. 
After constructing battle timelines of a match, a group of users, who engaged in a battle, has been extracted for each battle. 

Second, for each battle, the detection method evaluated logs of each participant in a battle, which includes the information of other users received from a server. After each battle, the ballots (evaluation results) of each user are organized, and the server evaluates them and calculates the two scores for each user based on the consensus algorithm.

\subsection{Customization of consensus algorithm}
We customized the consensus algorithm considering the characteristics of FPS.

\begin{description}[style=sameline,leftmargin=*]
    \item [Real-time battle]
        Since the battles occur sporadically, only a subset of users participates in a battle, not all users. Therefore, when \textit{vote} is initiated, voters are limited to users who participated in a battle.

    \item [Constant changing of users]
        During a battle, a number of users participating in a battle is constantly changing, due to death in combat or simply leaving a battlefield. In other words, a number of voters changes continually. However, users are not able to evaluate the user who killed early and vice versa. Thus, we limited voting rights to the users who participated in a battle long enough, so the method in each user could evaluate others. In addition, we also changed \textit{target\_user}, who is being evaluated by others, from all users to users who are detected by the detection method.

    \item [Self voting]
        Limiting and changing voters could have a negative impact for some cases. One example is when one normal user combats with one opposing cheating user (\textit{1 vs 1}) . If the ratio between cheating and normal users is set to fifty-fifty, the \textbf{Ground truth}, decided by the majority, could be manipulated by cheating users. Considering this, a vote for oneself was excluded. In addition, to prevent the intentional lie of cheating users targeting normal users, the \textit{vote} only occurs when more than two users participated and \textbf{Dubious} score was updated when the majority of users agreed that the \textit{target\_user} is a cheater.
\end{description}

A customized algorithm for the FPS genre can be found on \autoref{algo2:customized_censensus_algorithm} in \autoref{apx:consensus_algorithm}.

\subsection{Experimental design}

In previous research, 28 matches have been played, and logs of each player during a match have been recorded in the dataset (\cite{Choi_Ko}). The RNN model had been trained and tested on randomly divided logs of matches. We followed the process of previous research and split the logs based on match-basis (Note that it was named Game-based Split in the previous study). More specifically, this study performed \textit{k}-fold cross-validation, which uses \textit{k} - 1 sets of data for the training of a model and 1 set of data for the test of the trained model. In previous research, the \textit{k} was set to 7, since the divided sets could have the identical number of matches.

For each validation, we checked the final \textbf{Dubious} scores of users and determined a threshold that maximizes the \textit{Accuracy}, since game companies more worry about false alarms, such as \textbf{FN} and \textbf{FP}. (When a normal user is mistaken for a cheating user, and when a cheating user is mistaken for a normal user, respectively). Regarding the tactics of the liar, we adopted the \texttt{Random liar}, since it has proved to be a harsher condition for \verb|SYNOPTICON| as shown on \autoref{fig_Contour}.

\begin{figure*}[h]
\includegraphics[width=\columnwidth]{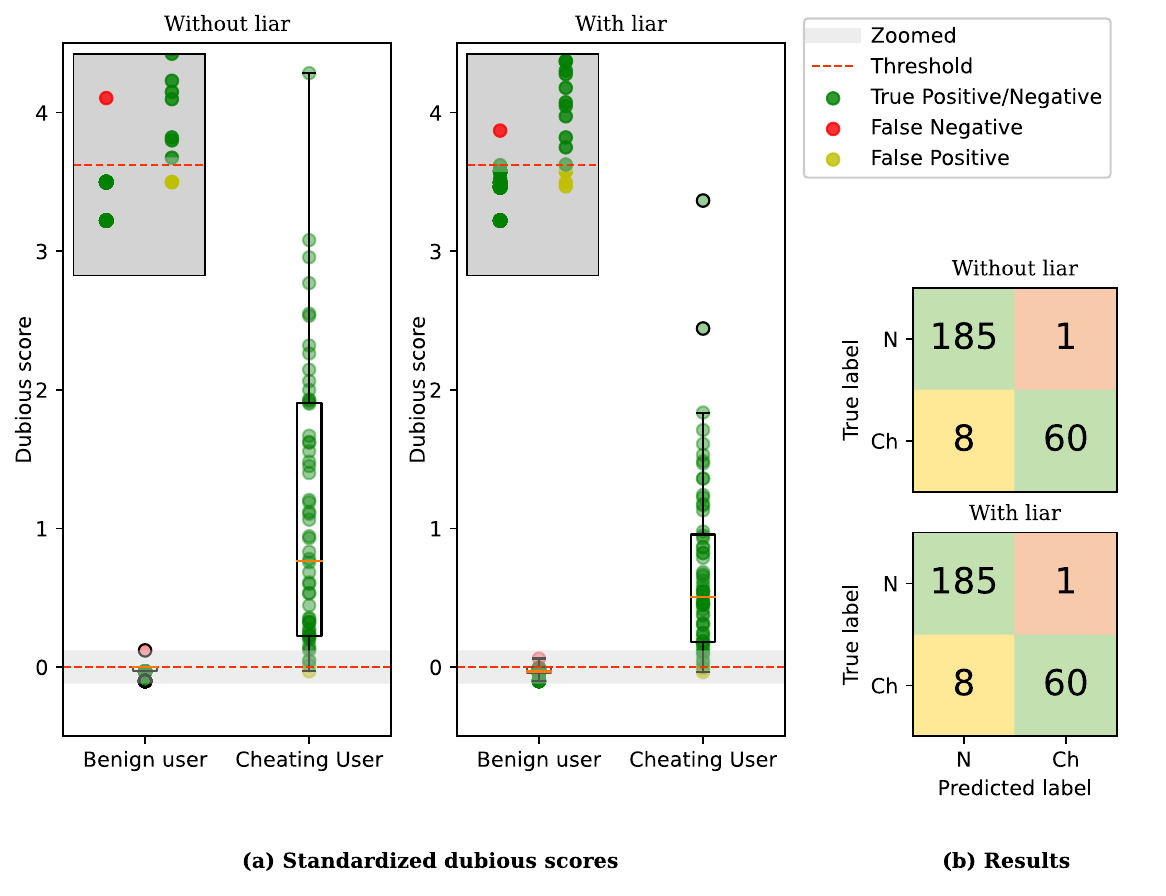}

\raggedleft \textmd{N: normal user Ch: cheating user} 
\captionof{figure}{Results of experiment. (a) demonstrates the standardized final \textbf{Dubious} scores of all validations and box plots of the final scores. To fix a threshold at 0 across all validations, for each validation, a threshold was subtracted from final scores. (b) is the confusion matrix of the results.}
\label{fig_gynopticon_result}
\Description[Results of experiment]{(a) demonstrates the standardized final Dubious scores of all validations and box plots of the final scores. To fix a threshold at 0 across all validations, for each validation, a threshold was subtracted from final scores. (b) is the confusion matrix of the results.}
\end{figure*}

\begin{table}[h]
\caption{Total number of votes for each match for which false alarms occured}
\label{tab3:case_study}
\begin{tabular}[\linewidth]{lrrlrr}
\toprule
\textbf{Result}             & \multicolumn{2}{c}{\textbf{Users}}                           &  & \multicolumn{2}{c}{\textbf{Votes}}                                \\ 
\cmidrule{2-3}\cmidrule{5-6}
                             & Total user & Cheater &  & Target user & Vote count  \\ 
\midrule
\textbf{FN}                  & 10                             & 4, 6                        &  & 9                               & 1                               \\ 
\midrule
\textbf{FP}                  & 8                              & 3, 4                        &  & 3                               & 0                               \\ 
\midrule
\textbf{FP}                  & 8                              & 2, 5                        &  & 5                               & 1                               \\ 
\midrule
\multirow{2}{*}{\textbf{FP}} & \multirow{2}{*}{8}             & \multirow{2}{*}{1, 4}       &  & 1                               & 0                               \\ 
\cmidrule{5-6}
                             &                                &                             &  & 4                               & 0                               \\ 
\midrule
\textbf{FP}                  & 10                             & 1, 9                        &  & 9                               & 0                               \\ 
\midrule
\textbf{FP}                  & 10                             & 0, 8                        &  & 0                               & 0                               \\ 
\midrule
\textbf{FP}                  & 10                             & 1, 5                        &  & 5                               & 0                               \\ 
\midrule
\textbf{FP}                  & 10                             & 1, 7                        &  & 1                               & 0                               \\
\bottomrule
\end{tabular}
\end{table}

\subsection{Results}\label{subsec:result}

\autoref{fig_gynopticon_result}-(a) shows standardized \textbf{Dubious} scores of all users for all validations, and \autoref{fig_gynopticon_result}-(b) demonstrates the overall experimental outcomes. We standardized the results of each validation by fixing a threshold at zero across all validations. Specifically, for each validation, we subtracted the determined threshold that maximizes the \textit{Accuracy} from the final \textbf{Dubious} scores to fix the threshold at zero across all validations. 

\autoref{fig_gynopticon_result}-(a) demonstrates that the tactics of liar worked to some extent, since the range of \textbf{Dubious} scores has been shorten. Still, as shown in \autoref{fig_gynopticon_result}-(b), \verb|SYNOPTICON| effectively differentiates normal users from cheating users, regardless of the presence of liar. However, the results appear to show some false alarms. To identify the underlying causes, we conducted in-depth analysis of the false cases and found that the primary cause of false alarms was a low \textit{vote count}. The \textit{vote count} for each false case is on \autoref{tab3:case_study}. The table shows that the \textit{vote} rarely occurred and fewer than one \textit{vote} was performed during a match, whereas an average \textit{vote count} for cheating users was five. Considering the results of the simulation on \autoref{fig_scatter} and \autoref{fig_Contour} in \autoref{sec:Methodoloy}, limited number of \textit{vote} and voters can negatively impact the stability of results. 

On the other hand, the reasons for the small \textit{vote count} were various. In most \textbf{FP} cases, because of an early death and leaving battlefield, cheating users participated in a battle too shortly, so other users can't evaluate them properly. Furthermore, limited number of large-scale battle, in which many users participated, contributed to the low \textit{vote count} and wrong evaluation. On the other hand, in \textbf{FN} cases, the primary issue was the accuracy of the detection method.

\section{Application and discussion} \label{sec:real_case_study_entry_game}

Even though \verb|SYNOPTICON| has been demonstrated its usability through simulation and its effectiveness with a real game dataset, the experiments mentioned above are limited to a game-play and also performed on a match-basis. 
Moving forward, for \textbf{Ch4}, we need to address the challenge of collecting reliable datasets from reliable sources. In addition, despite its usability, \verb|SYNOPTICON| should solve the ``50\% problem" of which the ratio of cheating users is more than 50\%, highlighted by the results on \autoref{fig_Contour}. We discuss these issues in this section. We carried out an additional simulation with our virtual game simulator based on a real-like environment. We implemented some operational policies and observed how we could distinguish cheating users from normal users to obtain reliable data. Finally, this section shows how well-developed policies can provide solutions to the seemingly unsolvable 50\% problem.

\subsection{Application on a real-like environment}

We simulated two types of competitive games of \textit{N vs N} and \textit{1 vs 1}. The first game represents typical competitive games and their settings, such as MOBA. For example, many of MOBA games follow the \textit{5 vs 5} setting including the `League of Legends' of Riot Games and `Dota2' of Valve. The second game is for other genres, such as Action and RTS. For example, many action games follow the \textit{1 vs 1} setting, such as `Tekken'of Bandai Namco and `Street Fighter' of CAPCOM. It is worth noting that the settings of the second game type could give hints about how to overcome the inherent limitation of \verb|SYNOPTICON| on which 50\% of users are cheating users; because, in \textit{1 vs 1} games only one \textbf{\textit{rigged vote}} by one cheating user can hold a majority and makes it hard to set the \textbf{Ground truth}.

\subsubsection{Simulation description}
\label{subsec:simulation_setting}

\begin{figure*}[htbp]
\includegraphics[width=\textwidth]{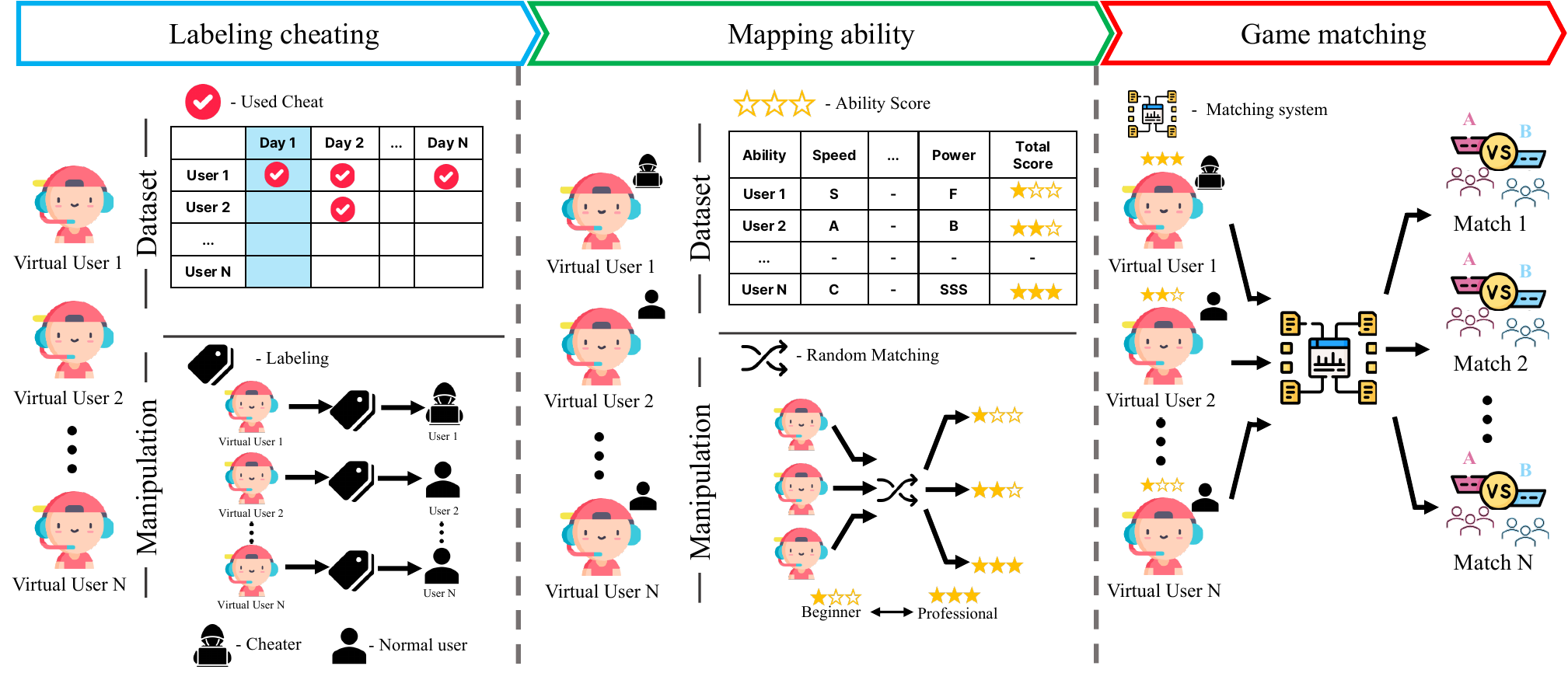}
\captionof{figure}{Simulation setting for a real-like online game. A detailed explanation and the dataset used for each step can be found in \autoref{subsec:simulation_setting}}
\label{fig_simulation_overall}
\Description[Simulation setting for a real-like online game]{A detailed explanation and the dataset used for each step can be found in \autoref{subsec:simulation_setting}}
\end{figure*}

In order to make our virtual game simulator as realistic as possible, which reflects real-world conditions, it was designed considering two factors: changing number of cheating users and the characteristics of a realistic online competitive game service. For the first factor, it was necessary to consider users' cheating usage patterns; because, several external elements in the real-world influence the users' cheating usage patterns. For example, one day, a normal user could start to use cheating, by watching one of their friends is using cheating or by yielding to the temptation of using cheating. Second factor is to reflect the characteristics of competitive games in the real-world. Competitive games are played on match-basis, and each match is a competition between two groups of users (or two users) with similar game-play abilities. To broaden our discussion from individual matches to overall game service and policy establishment, a realistic match-making system imitating a real-world game is essential. 

To satisfy requirements, we integrated two datasets from studies by \cite{deng2021globally} and \cite{woo2018contagion} into our simulation. Each steps for the integration is illustrated on \autoref{fig_simulation_overall}. In the first step, we created virtual users and each of them is labeled with the cheating usage pattern of the real-world users. The dataset was initially   published by \cite{woo2018contagion}. Their study examined the social contagion of cheating behavior found in social networks of online games by using bot detection logs of the real-world game named `Aion' developed by NCSoft. Because this dataset reflects real-world factors influencing users' cheating usage pattern, our simulation can help policymakers develop operational policies that can be applied in real-world scenarios.
Second, to accurately simulate a real-world game environment, we incorporated a match-making system and an ability dataset of real-world users from the study by \cite{deng2021globally}. The virtual users were randomly assigned ability scores, and matches were organized using the match-making system. Their study proposed a match-making system with efficiency and high performance. Most importantly, their method had been applied to the real-world games named `Fever Basketball' and `Justice Online' by Netease.

Our system simulates matches on a daily basis with ten thousand of virtual players participating in a certain number of matches per day, from one to nine. For the simulation of the first game with \textit{N vs N} setting, the number of \textit{N} was set to 3 and each user's \textbf{Validity} and \textbf{Dubious} scores were randomly matched with one of the scores in \autoref{subsec:result} after each match, according to the label attached at the first step, whether a user is a cheater or not. Since the user ability dataset only contains one thousand users' data, we employed over-sampling by randomly selecting and averaging the abilities of two users. The second game, simulating \textit{1 vs 1} games, follows the same setup as the first game, except that the simulator used in \autoref{sec:Methodoloy} is introduced to access users' two scores after each match. In a similar flow of \autoref{sec:Methodoloy}, we set the variables: \textit{benign num}(1), \textit{cheater num}(1), \textit{model acc}(80\%), and \textit{vote count} has been set randomly between 0 to 5. For the same reason in previous section, we adopted the \texttt{Random liar} as a tactic.

\subsubsection{Policy settings and implications}
For both games, we applied a policy for user evaluation and studied the results. Afterwards, we then explored potential policies that could be implemented to detect cheating users.

\begin{figure}[h!]
\includegraphics[width=\columnwidth]{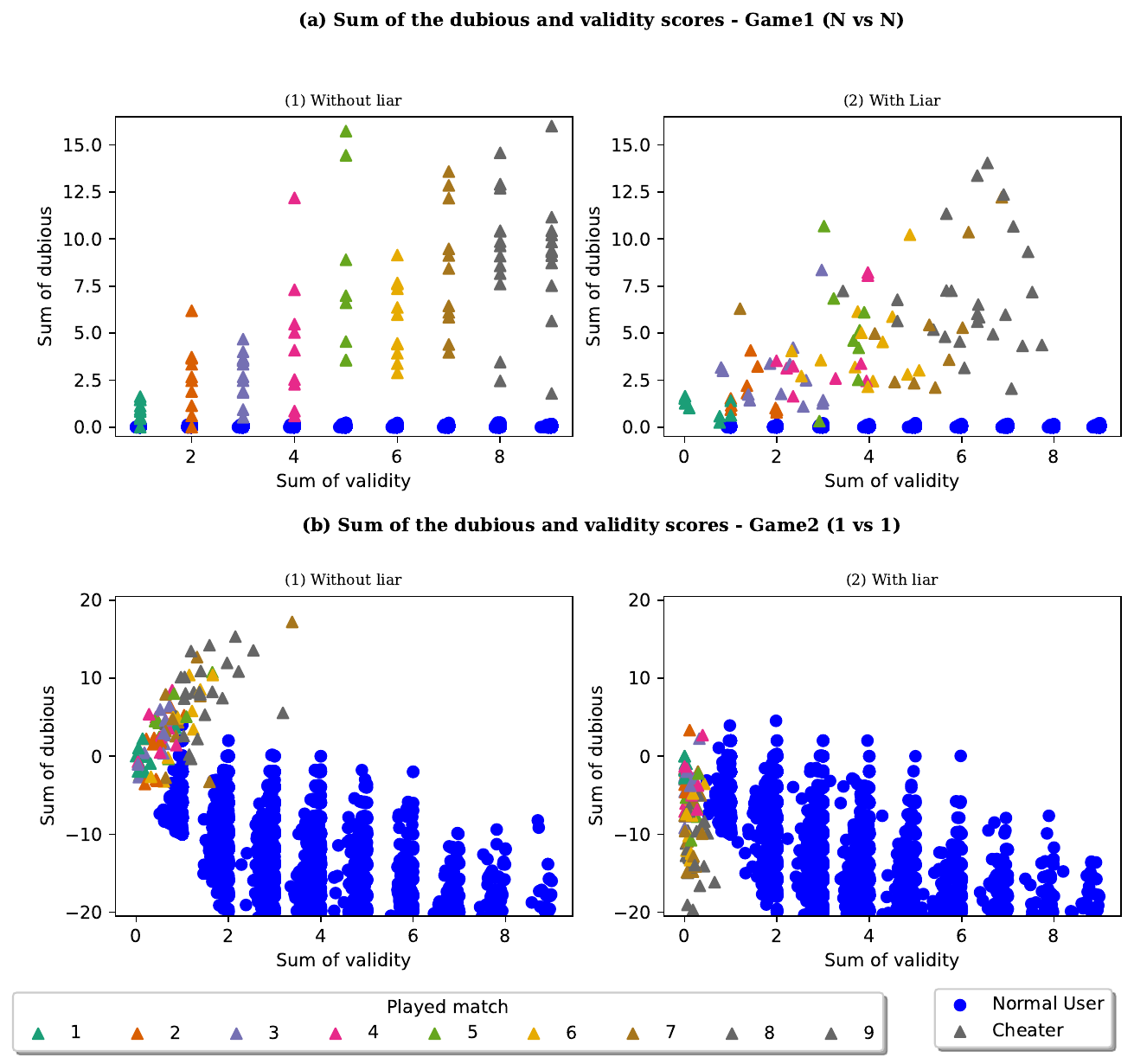}
\captionof{figure}{Result of the first policy. Users' \textbf{Dubious} and \textbf{Validity} scores have been added. \textbf{(a)} demonstrates the results of Game1 \textit{(N vs N)}. \textbf{(b)} demonstrates the results of Game2 \textit{(1 vs 1)}}
\label{fig_result_of_game1}
\Description[Result of the first policy]{Users' Dubious and Validity scores have been added. (a) demonstrates the results of Game1 (N vs N). (b) demonstrates the results of Game2 (1 vs 1)}
\end{figure}

\paragraph{Policy of adding.}

\begin{figure}[h!]
\includegraphics[width=\columnwidth]{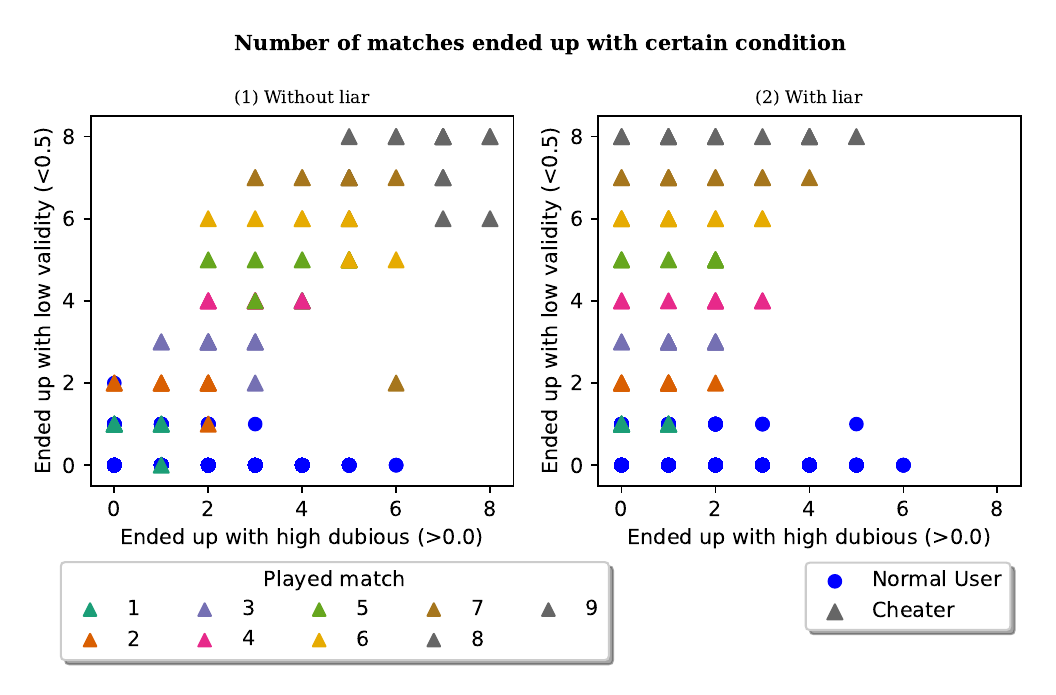}
\captionof{figure}{Result of the second policy implemented on Game2 \textit{(1 vs 1)}. The number of matches has been counted when a user satisfied specific conditions. The conditions are low \textbf{Validity} score (lower than 0.5) and high \textbf{Dubious} score (higher than 0.0)}
\label{fig_result_of_game2}
\Description[Result of the second policy implemented on Game2 (1 vs 1)]{The number of matches has been counted when a user satisfied specific conditions. The conditions are low Validity score (lower than 0.5) and high Dubious score (higher than 0.0)}
\end{figure}

For first policy, the evaluation proceeds by calculating the sum of the two scores each, after matches. 
\autoref{fig_result_of_game1} presents the overall scores after a one-day simulation. \autoref{fig_result_of_game1}-(a) demonstrates that even the simple policy of adding scores can effectively differentiate cheating and normal users based on the sum of the \textbf{Dubious} score.
However, \autoref{fig_result_of_game1}-(b) indicates that simply adding scores does not work efficiently for the second game. 

\paragraph{Policy of counting.}

We addressed the problem identified on \autoref{fig_result_of_game1}-(b) by leveraging the power of the majority. That is to say, if our consensus algorithm functions correctly, two normal users playing a match will both have high \textbf{Validity} and low \textbf{Dubious} scores. In contrast, when a normal user faces a cheater, two scenarios may occur: If the cheating user lies, both users will have low \textbf{Validity} and high \textbf{Dubious} scores. If the cheating user is honest, the cheating user will have high \textbf{Validity} and high \textbf{Dubious} scores, while a normal user will have high \textbf{Validity} and low \textbf{Dubious} scores. 

Based on this understanding, two players will have low \textbf{Validity} scores, if someone lies, and cheating users will have high \textbf{Dubious} scores if they don't lie. Therefore, we counted the number of matches when users satisfied certain conditions: when it has high \textbf{Dubious} scores and low \textbf{Validity} scores each. As a result, cheating users will be counted more frequently, and counted number of normal users will be distributed until the normal users take a majority in all users. The result on \autoref{fig_result_of_game2} demonstrates that cheating users can be filtered much clearly even in the \textit{1 vs 1} games. However, this strategy takes effect as far as normal users secure a majority in overall user-pool. 

\subsection{Discussion on applicability}

As previous sections present, implementing effective strategies can provide solutions for sustainable operation \textbf{(Ch4)} and help policymakers ensure the effective use of \verb|SYNOPTICON| \textbf{(Ch3)}---across various scenarios, such as \textit{1 vs 1}. The policymakers can collect reliable data from users to re-adjust  detection methods and quickly ban cheating users.
However, the two achievements require rapid detection speed \textbf{(Ch1)}. Let us say, if it takes one month, for example, policymakers may struggle to obtain reliable data from users; because, the data could be adulterated if one normal user starts to use cheating during the collecting. Consequently, a delayed ban may frustrate not only the policymakers but also users.

Moreover, more sophisticated policies may be needed. For instance, different detection methods may have to be applied to different user-pools depending on average abilities of users in each user-pool. If the gap of the abilities between professional and average players is huge, a detection method optimized for average users might mistakenly flag professional players as cheaters. In conclusion, despite the progress made in this section, we acknowledge that our approaches may be insufficient for some games. Therefore, in addition to \verb|SYNOPTICON|, we open our datasets and code, so others can test their advanced detection method and policies on our simulator. The sources are \href{https://anonymous.4open.science/r/Vega-E8D0}{available here}.

\section{Related work}
\begin{description}[style=sameline,leftmargin=*]
\item[Server-side \textit{aimbots} detection based on user behavior.]

Previous studies focusing on \textit{aimbots} adopted various methods for cheating detection. \cite{Liu_Gao_Zhang_Wang_Stavrou_2017} and \cite{Alayed_Frangoudes_Neuman_2013} differentiate cheating users with Support Vector Machine (SVM) classifiers. 
\cite{Galli_Loiacono_Cardamone_Lanzi_2011} compared five machine learning methods for detection---including Decision trees, Naive Bayes, Random forests, Neural
networks and SVM---and compared the results of detection. 
They evaluated the playing-log stacked on server-side received from client-side with statistics analysis and machine learning methods, yet server-side approaches are suffered from high overhead. 

\item[Client-side \textit{aimbots} detection based on user behavior.]

\cite{Spijkerman_Marie_Ehlers_2020a} conducted case studies of 9 experiments by comparing the results applied 3 detection methods (Decision Tree, SVM, Naive Bayes) to 3 datasets from different sources (Mouse input, Keyboard input, Game events), and provide valuable insights for choosing of detection method. In addition, \cite{pinto2021deep} used a machine learning method and evaluated keyboard-, mouse- inputs. Likewise, \cite{Choi_Ko} applied their RNN model for detection, and they used the character's coordinate and aiming angle as input data. Even though their client-side approaches achieved impressive results, the problem of low applicability is to be solved. However, \verb|SYNOPTICON| can be integrated with their prestigious research and will solve the problem.

\item[Other approaches for cheating detection.]
As different approaches, \cite{Han_Park_Kim_2015} introduced statistics analysis. They use various features for detection, such as winning rates and play-time. \cite{Yu_Hammerla_Yan_Andras_2012a,Yu_Hammerla_Yan_Andras_2012b} assessed user input, such as cursor movement and cursor acceleration. The statistical approach may not perfectly fit with \verb|SYNOPTICON|, as it takes a relatively long time and is proven less accurate. Still, in the operational aspect, it would give meaningful insights for policymakers. To achieve this, our real-like simulation can be applied. \cite{dong2018gci} and \cite{islam2020gci} adopted network-based approach. They analyzed encrypted network traffic by using the machine learning method. Even though their approach may differ from approaches based on user behavior, network-based approaches can also be applied to \verb|SYNOPTICON| as well; because, our basic assumption is based on using data received from a server.

\end{description}

\section{Conclusion and Perspective}
Cheating detection in competitive games got less attention from the academy and industry as well. Even for its increasing size in the gaming market, it was less attractive for researchers, as it requires sophisticated detecting methods and datasets. Existing studies, also, suffered from several challenges including low applicability and high overhead. In this work, we addressed those challenges that have been faced in previous studies. At first, we discussed that applying a detecting method on server-side could incur a considerable cost of overhead, yet if we apply the method on client-side, there is also a chance of tampering by clients, which is unrealistic to apply. Anti-cheat solutions, which is commonly used in the industry, also have problems in terms of user privacy. Next, we introduced \verb|SYNOPTICON|. \verb|SYNOPTICON| assumes that the detection method was applied to all users, so server-side can keep overhead low. Tampering by users is protected by normal users' consensus. Lastly, \verb|SYNOPTICON| does not require personal information of users, as it only uses data received from server-side. In discussion, we practiced that \verb|SYNOPTICON| can be applied to real-world games and showed that it successfully worked as a cheating detection system. 

Future research on cheating detection should focus on more various competitive genres, such as MOBA and Action, as cheating detection methods regarding these are not addressed yet, to the best of our knowledge. Furthermore, while our experiments proved the effectiveness of policy setting, studies on cheating detection methods of competitive games are required to offer more insight into operational aspects  of their method. Especially on competitive games, discussions about operation and managing are insufficient.

\bibliographystyle{ACM-Reference-Format}
\bibliography{references}

\appendix

\section{Consensus algorithm} \label{apx:consensus_algorithm}
\subsection{Parameter selection}
The algorithm has been designed very carefully to consider both \textbf{\textit{error vote}} and \textbf{\textit{rigged vote}}. The discussion about \textbf{\textit{rigged vote}} was already held in the paper, so we skip its explanation now. In the case of \textbf{\textit{error vote}}, the history queue ($Q_N$) has been introduced to prevent decreasing in \textbf{Validity} score of normal users by \textbf{\textit{error vote}}. To give weights on the user's ballot and history, two variables of $W_v$ and $W_h$ have been set. In our design, $W_v$ and $W_h$ are set to $0.05$ and $0.01$ each. The values only affect the final scores, so we have to discuss the rate between the two values. $W_h$ is set to one-fifth of $W_v$, and this setting performs ideally when the accuracy of the detection method is 80\%. As an illustration, for normal users, it is reasonable to expect that they would have two \textbf{\textit{error vote}} in the history queue on average, since the maximum size of the queue is ten $(10 \times \frac{(100-80)}{100}\%$). Therefore, the ratio of true and false ballots in the queue might be 8 : 2. In this case, even if they vote \textbf{\textit{error vote}} once more, it will not affect their \textbf{Validity} scores. However, if someone starts to say lies over previous assumption, it will start to lose its influence on vote.

\subsection{Customized algorithm}
In addition to the consideration in the previous section, the details of the algorithm can be found on  \autoref{algo2:customized_censensus_algorithm}.

\begin{algorithm}
\SetKwData{True}{True}
\SetKwData{False}{False}
\SetKwFunction{Count}{count}
\SetKwFunction{Enque}{enque}
\SetKwFunction{Deque}{deque}
\SetKwFunction{Pop}{pop}
\SetKwFunction{Length}{length}

\SetKwInput{GT}{\textit{GT}}
\SetKwInput{Users}{\textit{Users}}
\SetKwInput{VadW}{$W_v$}
\SetKwInput{HisW}{$W_h$}
\SetKwInput{History}{$Q_N$}
\SetKwInput{Dubious}{$D_N$}
\SetKwInput{Validity}{$V_N$}
\SetKwInput{UserV}{$Vote_{target}(N)$}
\SetKwFunction{FMain}{score\_evaluation}
\SetKwFunction{FDoVote}{do\_vote}
\SetKwProg{Fn}{Function}{:}{}
\SetKwComment{comment}{}{} 

 \GT {Ground truth}
 \Users {List of users}
 \VadW {Weight for Validity}
 \HisW {Weight for History}
 \History {Queue for voting record of user\textit{N}}
 \Dubious {Dubious score of user\textit{N}}
 \Validity {Validity score of user\textit{N} $(0 < V_N < 1)$} 
 \UserV {A vote of user\textit{N} for \textit{target}}
 
  \Fn{\FMain{$target\_user$}}{
        \comment{\ding{192}-Count users' votes and set ground truth }
        
        $c\_cnt \leftarrow 0$ \comment*[r]{ is Cheater }
        $n\_cnt \leftarrow 0$ \comment*[r]{ is Normal user }
      \ForEach{$user$ in the \textbf{Users}}{
          \lIf(\comment*[f]{\ding{193}-Skip self voting}){$user$ is $target\_user$}{
                continue
            }
            \lIf{$Vote_{target\_user}(user)$ is \True}{
                $c\_cnt \leftarrow c\_cnt + 1$
            }
            \lElseIf{$Vote_{target\_user}(user)$ is \False}{
                $n\_cnt \leftarrow n\_cnt + 1$
            }
            \lElse(\comment*[f]{Pass users who didn't vote}){
                continue
            }
      }
      \lIf{$c\_cnt > n\_cnt$}{
            \textbf{GT} $\leftarrow$ \True
        }
        \lElse{
            \textbf{GT} $\leftarrow$ \False
        }
    \ForEach{$user$ in the \textbf{Users}}{
          \lIf(\comment*[f]{\ding{193}-Skip self voting}){$user$ is $target\_user$}{
            continue
        }

        $t\_cnt \leftarrow Q_{user}.$\Count{\True} \;
        $f\_cnt \leftarrow Q_{user}.$\Count{\False} \;
        
        $hist\_record \leftarrow t\_cnt - f\_cnt$\;
        \lIf{$Q_N.\Length{}>10$}{$Q_N.$\Deque{}}
        
        \lIf{\begin{math}Vote_{target\_user}(user)\ne\textbf{GT}\end{math}}
        {user\_lied $\leftarrow$\True}
        \lElse{user\_lied $\leftarrow$\False}
        
        \uIf{user\_lied is \True}{
            \begin{math}V_{user} \leftarrow V_{user}-W_v + hist\_record*W_h\end{math}\;
            $Q_{user}.$\Enque{\False}\;
        }
        \Else{
            \begin{math}V_{user} \leftarrow V_{user}+W_v+hist\_record*W_h\end{math}\;
            $Q_{user}.$\Enque{\True}\;
        }
        \comment{\ding{194}-Calculate Dubious when majority agreed }
        \If{$c\_cnt$ $>$ $n\_cnt$ and $c\_cnt + n\_cnt > 2$}{
            \comment{\ding{195}-Restrict voters'}
            \uIf{$Vote_{target\_user}(user)$ is \True}{
                \begin{math}D_{target\_user} \leftarrow D_{target\_user}+\dfrac{c\_cnt}{c\_cnt+n\_cnt}V_{user}\end{math}\;
            } \uElse{
                \begin{math}D_{target\_user} \leftarrow D_{target\_user}-\dfrac{n\_cnt}{c\_cnt+n\_cnt}V_{user}\end{math}\;
            }
        }
    } 
  }

\comment{\ding{196}-Vote only for \textit{target\_user}}
\Fn{\FDoVote{$target\_user$, \textbf{GT}}}{
    \FMain{$target\_user$, \textbf{GT}}
}
  
\caption{Customized version}
\label{algo2:customized_censensus_algorithm}

\end{algorithm}

\begin{enumerate}
    \itemindent=-9pt
    \item[\ding{192}]
    Limit voters to participant in battlefield ($c\_cnt$, $n\_cnt$). When choosing the \textbf{Ground truth}, we don't consider users who didn't participate to \textit{vote}.
    \itemindent=-9pt
   \item[\ding{193}] 
    We don't consider the \textbf{Self voting} into account. 
    \itemindent=-9pt
   \item[\ding{194}] 
    The \textbf{Dubious} score is updated when the majority of users agree that the \textit{target\_user} is a cheating user.

    \itemindent=-9pt
   \item[\ding{195}]
    Participated users have the same influence, by dividing $c\_cnt + n\_cnt$, and majority voting have more power, by multiplying $c\_cnt$ and $n\_cnt$.
    \itemindent=-9pt
   \item[\ding{196}]  
   A \textit{vote} occurs only for a \textit{target\_user}.
 \end{enumerate}

\end{document}